\documentstyle[11pt]{article}

\def\singlespaced{\baselineskip=\normalbaselineskip}    
\def\singlespace{\singlespaced}                         

\def\widenspacing{\multiply\baselineskip by 150         
    \divide\baselineskip by 100}                        
\def\whitespace{\widenspacing}                          

\newcommand{\vph}{\varphi}

\newcommand{\ext}{\hbox{\it ext}}
\newcommand{\comp}{\hbox{\scriptsize\it COMP}}
\newcommand{\tree}{\hbox{\it tree}}
\newcommand{\cwa}{\hbox{\scriptsize\it CWA}}
\newcommand{\Cwa}{\hbox{\it CWA}}

\newcommand{\at}{\hbox{\it At}}

\newcommand{\la}{\langle}
\newcommand{\ra}{\rangle}

\newtheorem{theorem}{Theorem}[section]

\newtheorem{corollary}[theorem]{Corollary}
\newtheorem{example}{Example}[section]
\newtheorem{definition}{Definition}[section]
\newtheorem{proposition}[theorem]{Proposition}

\newtheorem{remark}{Remark}[section]
\newtheorem{theorem1}{Theorem}[subsection]

\newtheorem{example1}{Example}[subsection]
\newtheorem{definition1}{Definition}[subsection]

\newtheorem{remark1}[theorem1]{Remark}

\newenvironment{bdefinition}{\begin{definition} \rm}{\end{definition}}

\begin{document}
\ \\
\begin{center}
{\Large\bf Representation Theory for Default Logic}
\ \\
\ \\
{\it V. Wiktor Marek\footnote{Department of Computer Science,
University of Kentucky, Lexington, KY 40506-0046, USA. E-mail:
{\tt marek@cs.engr.uky.edu}, Fax: +(606) 323-1971.}
     Jan Treur\footnote{Free University Amsterdam, Department of Mathematics 
              and Computer Science, Artificial Intelligence Group,
              De Boelelaan 1081a, 1081 HV  Amsterdam, The Netherlands,
{\tt treur@cs.vu.nl}.} and
     Miros\l aw Truszczy\'nski\footnote{Department of Computer Science,
University of Kentucky, Lexington, KY 40506-0046, USA, 
{\tt mirek@cs.engr.uky.edu},  Fax: +(606) 323-1971.}}
\end{center}
\begin{center}
{\Large\bf Abstract}
\end{center}

{\small
\noindent
Default logic can be regarded as a mechanism to represent
{\em families} of belief sets of a reasoning agent. As such, it is
inherently second-order. 
In this paper, we study the problem of representability of a family of
theories as the set of extensions of a default theory. We give a
complete solution to the representability by means of normal default
theories. We obtain partial results on representability by arbitrary
default theories. We construct examples of denumerable
families of non-including theories that are not representable.
We also study the concept of equivalence between default theories.   
}
\ \\
\ \\
\newpage
\whitespace
\section{Introduction}

In this paper we investigate the issues related to the expressibility of
default logic, a knowledge representation formalism introduced by
Reiter \cite{re80} and extensively investigated by the
researchers of logical foundations of Artificial Intelligence
\cite{et88,bes89,brb91}. A default
theory $\Delta$ describes a family (possibly empty) 
of belief sets of an agent reasoning
with $\Delta$. In that, default logic is inherently second-order, but in a sense
different from that used by logicians. Whereas a logical theory  $S$
describes a subset of the set of all formulas (specifically, the set 
$Cn(S)$ of logical  consequences of $S$), a default theory 
$\Delta$ describes a collection of subsets of the set of all formulas, 
namely the family of all extensions
of $\Delta$, $\ext(\Delta)$. Hence, default theories can be viewed as 
encodings of families of subsets of some universe described 
by a propositional language. Examples of encodings of the problems of
existence of colorings, kernels, and hamilton cycles in graphs
are given in \cite{cmmt94}.

This second-order flavor of default logic makes it especially
useful in knowledge representation. An important question, then,
is to characterize those families of sets that can be represented as 
the set of extensions of a certain default theory. This is the topic of
our paper.

There is a constraint 
on the family $\cal T$ of extensions of a default theory $\Delta$. Namely
such family must be {\em non-including} \cite{re80}. In this paper we 
exhibit several classes of families of non-including theories that
can be represented by default theories. We also show that there are
non-representable families of non-including theories.
The existential proof follows easily from a cardinality 
argument. There are continuum-many default theories in a given 
(denumerable) language, while there is more than continuum-many families of 
non-including theories. In the paper, we actually {\em construct} 
a family of non-including theories that is not represented by 
a default theory. Moreover, our family is 
denumerable (the cardinality argument mentioned above does not guarantee
the existence of a non-representable denumerable family of non-including
theories).

The family of extensions of a {\em normal} default theory is not only
non-including, but all its members are pairwise inconsistent \cite{re80}. 
In this paper, we fully characterize these families of theories
which are of the form $\ext (\Delta)$, for a normal default theory  $\Delta$.
In addition, we construct examples of denumerable families of pairwise 
inconsistent  theories which are not representable by normal default
theories.

Let ${\cal T} = \ext(\Delta)$, for some default theory $\Delta$.
Clearly, there are other default theories $\Delta'$ such that
$\ext(\Delta')= {\cal T}$. In other words, $\Delta$ is not uniquely
determined by $\cal T$. Thus, it is natural to search
for alternative default theories $\Delta'$ with the same set of extensions 
as $\Delta$. Let us call  $\Delta'$ equivalent to $\Delta$ if 
$\ext (\Delta) = \ext (\Delta')$. We show that for every $\Delta$ we can 
effectively (without constructing extensions of $\Delta$) find an equivalent 
theory $\Delta'$ with all defaults in $D'$ prerequisite-free (this
result was obtained independently by Schaub \cite{schau92a}, and Bonatti
and Eiter \cite{be94}). An  important feature of our approach
is that it shows that
when $\Delta$ is normal, we can construct a {\em normal} prerequisite-free 
default theory $\Delta'$ equivalent to $\Delta$.

We also present results that allow us to replace some normal theories $\Delta$
with equivalent normal default theories of the form $(D',\emptyset)$. 
At present, it is an open problem to decide whether such replacement 
is possible for every normal default theory $\Delta$ with $W$ consistent.

We discuss yet another (weaker) form of equivalence and prove that every
normal default theory is equivalent to a theory closely related to the
closed world assumption over a certain set of atoms.

This paper sheds some light on the issue of expressibility of default logic 
and, in particular, on expressibility of normal default logic.
We firmly believe that the success of default logic 
as a knowledge representation mechanism depends on a deeper understanding
of expressibility issues.

\section{Preliminaries}

In this paper, by $\cal L$ we denote a language of propositional logic with a denumerable set of atoms $\at$. By a {\em theory} we always mean 
a subset of $\cal L$ {\em closed under propositional provability}. 
Let $B$ be a set of standard monotone inference rules. The formal system
obtained by extending propositional calculus with the rules from $B$ will be denoted
by $PC+B$. The corresponding provability operator will be denoted by $\vdash_B$
and the consequence operator by $Cn^B(\cdot)$ \cite{mt93}.

A {\em default} is an expression $d$ of the form
$\frac{\alpha\colon\Gamma}{\beta}$, where $\alpha$ and $\beta$ are formulas from
$\cal L$ and $\Gamma$ is a {\bf finite} subset of $\cal L$. The formula $\alpha$
is called the {\em prerequisite}, formulas in $\Gamma$ --- the {\em justifications},
and $\beta$ --- the {\em consequent} of $d$. The prerequisite, the set of justifications and 
the consequent of a default $d$ are denoted by $p(d)$, $j(d)$ and $c(d)$, respectively.
If $p(d)$ is a tautology, $d$ is called {\em prerequisite-free} ($p(d)$ is then usually omitted from the notation of $d$).
This terminology is naturally extended to a set of defaults $D$.

By a {\em default theory} we mean a pair $\Delta =(D,W)$, where $D$ is 
a set of defaults and $W$ is a set of formulas, is called a {\em default 
theory}.  A default theory $\Delta = (D,W)$ is called {\em finite} if both 
$D$ and $W$ are {\em finite}.
For a set of defaults $D$, define 
\[
Mon(D) = \left\{\frac{p(d)}{c(d)}\colon d\in D\right\}.
\]
A default $d$ (a set of defaults $D$) is {\em applicable} with respect 
to a theory $S$ (is {\em$S$-applicable}) if $S\not\vdash\neg \gamma$ 
for every $\gamma\in j(d)$ 
($j(D)$, respectively). Let $D$ be a set of defaults. By the {\em reduct} 
$D_S$ of $D$
with respect to $S$ we mean the set of monotone inference rules:
\[
D_S= Mon(\{d\in D\colon \mbox{$d$ is $S$-applicable}\}).
 \]
A theory $S$ is an {\em extension}\footnote{Our definition of extension is different from but equivalent to the original definition by Reiter. See \cite{mt93} for details.}
of a default theory $(D,W)$ if and only if 
\[
S = Cn^{D_S}(W).
\]
The family of all extensions of $(D,W)$ is denoted by $\ext(D,W)$.

A family $\cal T$ of subsets of $\cal L$ is {\em non-including} if:
\begin{enumerate}
\item each $T\in {\cal T}$ is closed under propositional consequence, and
\item $\cal T$ is an antichain, that is, for every $T,T'\in {\cal T}$,
if $T\subseteq T'$ then $T=T'$.
\end{enumerate} 

Let $S$ be a theory. A default $d$ is {\em generating} for $S$ if $d$ is 
$S$-applicable and  $p(d)\in S$. The set of all defaults in $D$ generating 
for $S$ is denoted by $GD(D,S)$. It is well-known \cite{mt93} that
\begin{description}
\item[(P1)] If $S$ is an extension of $(D,W)$ then $S = Cn(W\cup c(GD(D,S)))$,
\item[(P2)] If all defaults in $D$ are prerequisite-free then $S$ is an extension of $(D,W)$
if and only if $S = Cn(W\cup c(GD(D,S)))$.
\end{description}

We will define now the key concepts of the paper.

\begin{bdefinition}\label{equi}
Default theories $\Delta$ and $\Delta'$ are {\em equivalent} if
$\ext(\Delta) = \ext(\Delta')$.
\end{bdefinition}

\begin{bdefinition}\label{s-equi}
Let $\Delta$ be a default theory over a language $\cal L$ and let $\Delta'$ be 
a default theory over a language ${\cal L}'$ such that ${\cal
L}\subseteq {\cal L}'$. Theory $\Delta$ is semi-equivalent to $\Delta'$
if $\ext(\Delta) = \{T\cap {\cal L}\colon T\in \ext(\Delta')\}$.
\end{bdefinition}

\begin{bdefinition}\label{rep.ft}
Let $\cal T$ be a family of  theories contained in
$\cal L$. The family $\cal T$ is {\em representable} by a default 
theory $\Delta$ if $\ext(\Delta)={\cal T}$.
\end{bdefinition}

\section{Default theories without normality restriction}

We start with the result that allows us to replace any default theory with an
equivalent  default theory
in which all defaults are prerequisite-free. As mentioned, this result was known before.
However, our argument shows that if we start with a normal default theory, 
its prerequisite-free equivalent replacement can also be chosen to be normal.

\begin{theorem}\label{prereq-free}
For every default theory $\Delta$ there is a prerequisite-free default
theory $\Delta'$ equivalent to $\Delta$. Moreover, if $\Delta$ is
normal
then $\Delta'$ can be chosen to be normal, too.
\end{theorem}
Proof: Let $\Delta=(D,W)$. By a {\em quasi-proof} from $D$ and $W$ we 
mean any proof from $W$ in the system $PC+Mon(D)$.
For every quasi-proof $\epsilon$ from $D$ and $W$ let
$D_\epsilon$ be the set of all defaults used in $\epsilon$.
For each such proof $\epsilon$, define
\[
d_\epsilon = \frac{\ \colon j(D_\epsilon)}{\bigwedge cons(D_\epsilon)}.
\]
Next, define
\[
E = \{d_\epsilon\colon \epsilon \ \mbox{is a quasi-proof from $W$}\}.
\]
Each default in $E$ is prerequisite-free.
Put $\Delta'=(E,W)$. We will show that 
$\Delta'$ has exactly the same extensions as $(D,W)$. To this 
end, we will show that
for every theory $S$ and for every formula $\vph$,
\[
W\vdash_{D_S}\vph \ \ \mbox{iff}\ \ W\vdash_{E_S}\vph .
\]
Assume first that $W\vdash_{D_S}\vph$. Then, there is a quasi-proof
$\epsilon$ of $\vph$ such that all defaults in $D_\epsilon$ are applicable
with respect to $S$. Moreover,  $W\cup c(D_\epsilon)\vdash\vph$.
Observe that $c(d_\epsilon) \vdash c(D_\epsilon)$. Since
$d_\epsilon$ is prerequisite-free and $S$-applicable, 
$W\vdash_{E_S} W\cup c(D_\epsilon)$.
Hence, $W\vdash_{E_S} \vph$.

To prove the converse implication, observe that since all defaults in 
$E$ are prerequisite-free, 
\[
\{\vph\colon W\vdash_{E_S}\vph \} = 
Cn(W\cup c(E_S)).
\]
Hence, it is enough to show that
\[
W\vdash_{D_S}  W\cup c(E_S).
\]
Clearly, for every $\vph\in W$,  $W\vdash_{D_S}\vph$. 
Consider then $\vph\in c(E_S)$. It follows
that there is a quasi-proof $\epsilon$ such that $d_\epsilon$ is $S$-applicable and $c(d_\epsilon) =\vph$. Consequently,
all defaults occurring in $\epsilon$ are $S$-applicable.
Thus, for every default $d\in D_\epsilon$, 
\[
W\vdash_{D_S}c(d).
\]
Since $\vph = \bigwedge c(D_\epsilon)$, 
\[
W\vdash_{D_S}\vph.
\]

To prove the claim for normal default theories, observe that
if each default in $D$ is normal, then each default in
$E$ is of the form
\[
\frac{\ \colon \Gamma}{\bigwedge \Gamma}.
\]
Let ${\widehat E}$ be a set of defaults obtained from 
$E$ by replacing each default
$\frac{\ \colon \Gamma}{\bigwedge \Gamma}$ by the normal default
$\frac{\ \colon \bigwedge\Gamma}{\bigwedge \Gamma}$. It is easy to
see that $S$ is an extension of $(E,W)$ if and only if
$S$ is an extension of $({\widehat E},W)$. \hfill$\Box$
\ \\

The next result fully characterizes families of theories representable by 
default theories with a finite set of defaults.

\begin{theorem} \label{th-1}
The following statements are equivalent:
\begin{enumerate}
\item[(i)] ${\cal T}$ is representable by a default theory $(D,W)$ with
finite $D$
\item[(ii)] ${\cal T}$ is a finite set of non-including theories, finitely 
generated over the intersection of $\cal T$
\end{enumerate}
\end{theorem}
Proof: Assume (i). Since every extension of $(D,W)$ is of the form
$Cn(W\cup c(D'))$, for some $D'\subseteq D$, it follows that
$\ext(D,W)$ is finite. It is also well-known (\cite{re80,mt93})
that $\ext(D,W)$ is non-including. Let $U$ be the intersection of all 
theories in 
$\ext(D,W)$. Then $W\subseteq U$. Consequently, each extension in
$\ext(D,W)$ is of the form $Cn(U\cup c(D'))$. Hence, each extension
is finitely generated over the intersection of $\ext(D,W)$.

Now, assume (ii). Let $U$ be the intersection of all theories in $\cal T$.
It follows that there is a positive  integer $k$ and formulas $\vph_1,\ldots,\vph_k$
such that ${\cal T}=\{T_1,\ldots,T_k\}$ and each 
$T_i = Cn(U\cup\{\vph_i\})$. 

Assume first that $k=1$. Then, it is evident that ${\cal T}$ is the family
of extensions of the default theory $(\emptyset, T_1)$. Hence,
assume that $k\geq 2$. Since the theories in $\cal T$ are non-including,
for every $j\not= i$ we have 
\begin{equation}\label{eq-2}
U\cup\{\vph_i\}\not\vdash \vph_j.
\end{equation}
In particular, each theory in $\cal T$ is consistent and so is $U$.
Moreover, it follows from (\ref{eq-2}) that for every $j=1,\ldots,k$,
\begin{equation}\label{eq-3}
U\not\vdash \vph_j.
\end{equation}
Define 
\[
d_i=\frac{\ \colon \{\neg \vph_1,\ldots,\neg\vph_{i-1},
                             \neg\vph_{i+1},\ldots,\neg\vph_k\}}{\vph_i},
\]
$i=1,\ldots,k$. Next, define $D= \{d_1,\ldots,d_k\}$. 
We will show that $\ext(D,U)={\cal T}$. 

Let $T$ be an extension of $(D,U)$. Then, there is a subset $\Phi$
of $\{\vph_1,\ldots,\vph_k\}$ such that $T=Cn(U\cup \Phi)$. Assume that
$|\Phi|=0$. Then, by (\ref{eq-3}), 
$D_T=\{\frac{\ :\ }{\vph_i}\colon i=1,\ldots,k\}$. 
Consequently, $U = T = Cn^{D_T}(U) = Cn(U\cup\{\vph_1,\ldots,\vph_k\})$. 
Hence, for every $i$, $U\vdash \vph_i$, a contradiction (with
(\ref{eq-3})). Hence, $|\Phi|>0$.
Assume that $|\Phi|>1$. By the definition of $D$, $D_T = \emptyset$.
Consequently, $T = Cn(U\cup\Phi) = Cn^{D_T}(U)=Cn(U)$. Let $\vph\in \Phi$
(recall that $\Phi\not=\emptyset$). Then, $U\vdash\vph$, a contradiction.
Hence, every extension $T$ of $(D,W)$  is of 
the form $Cn(U\cup\{\vph_i\})$ for some $i$, $1\leq i\leq k$.

To complete the proof, consider an arbitrary $i$, $1\leq i\leq k$.
We will show that $T_i$ is an extension of $(D,W)$. First,
observe that, by (\ref{eq-2}), $D_{T_i}= \{\frac{\ :\ }{\vph_i}\}$. 
Consequently, $Cn^{D_{T_i}}(U) = Cn(U\cup\{\vph_i\})=T_i$.
Hence, $T_i$ is an extension of $(D,U)$. \hfill$\Box$
\ \\

This result and its argument provide the following corollary which gives 
a complete characterization of families of theories representable by 
finite default theories, that is, theories $(D,W)$ with both
$D$ and $W$ finite.

\begin{corollary} \label{th-2}
The following statements are equivalent:
\begin{enumerate}
\item ${\cal T}$ is representable by a finite default theory
\item ${\cal T}$ is a finite set of finitely generated non-including theories
\end{enumerate}
\end{corollary}

%
%


As pointed out in the introduction, the cardinality argument implies
the existence of non-representable families of non-including theories.
However, it does not imply the existence of denumerable non-representable families.
We will now show two examples of such families. The first family consists of 
non-including finitely generated theories. The second one consists of 
mutually inconsistent theories.

\begin{theorem}\label{non-rep-1}
There exists a countable family of finitely generated
non-including theories  ${\cal T}$  such that  ${\cal T}$ is not 
representable by a default theory.
\end{theorem}
Proof: Let $\{p_0,p_1,\ldots\}$ be a set of propositional atoms.
Define $T_i=Cn(\{p_i\})$, $i=0,1,\ldots$, and ${\cal T}=\{T_i\colon
i=0,1,\ldots\}$. It is clear that $\cal T$ is countable and consists
of non-including theories. We will show that $\cal T$ is not 
representable by a default theory.

Assume that $\cal T$ is represented by a default theory $(D,W)$.
By Theorem \ref{prereq-free}, we may assume that all defaults in
$D$ are prerequisite-free. We can also assume that no default in $D$
contains a justification which is contradictory (such defaults are never
used to construct extensions). 

Consider a default $d\in D$. Since $j(d)$ is finite, there is $k$
such that for all $m>k$, all formulas in $j(d)$ are consistent
with $T_m$. Since $T_m$ is an extension of $(D,W)$, $c(d)\in T_m$,
for $m>k$. Since 
\[
\bigcap_{m>k} T_m = Cn(\emptyset),
\]
$c(d)$ is a tautology. Since $d$ was arbitrary, it follows that
$(D,W)$ possesses only one extension, namely $Cn(W)$, a contradiction.
\hfill$\Box$
\ \\

\begin{theorem}\label{non-rep-2}
There exists a countable family of  mutually inconsistent
theories  ${\cal T}$  such that  ${\cal T}$ is not 
representable by a  default theory. In particular ${\cal T}$ is not
representable by a  normal default theory.
\end{theorem}
Proof:  Let $\{p_0,p_1,\ldots\}$ be a set of propositional atoms.
Define $$T_i=Cn(\{\neg p_i,p_{i+1},\ldots\}),$$ for $i=0,1,\ldots$, 
and ${\cal T}=\{T_i\colon
i=0,1,\ldots\}$. It is clear that $\cal T$ is countable and consists
of pairwise inconsistent theories. Now, we apply precisely the same
argument as in the proof of Theorem \ref{non-rep-1}.
\hfill$\Box$
\ \\

Our counterexamples have an additional property that their infinite subsets
and all supersets are also counterexamples.

Finally, we show that if infinite sets of justifications are 
allowed in defaults,
every theory of non-including theories can be represented as the family
of extensions. 

\section{Eliminating extensions}

In this section, we consider the problem of representability of
subfamilies of a representable family. We present a technique for 
constructing default theories representing some subfamilies of 
a family of extensions of a given default theory $\Delta$. Such techniques 
are important when we have to redesign the default theory to exclude
extensions containing a specific formula and preserve all the remaining
extensions unchanged.

Let $\vph\in{\cal L}$. Define
\[
d_{\vph} = \frac{\vph:\ }{\bot}.
\]

\begin{theorem}\label{th.1}
Let $E\subseteq{\cal L}$ be consistent and let $(D,W)$ be a default
theory. Then, $E$ is an extension of $(D\cup\{d_{\vph}\},W)$ if and
only if $\vph\notin E$ and $E$ is an extension of $(D,W)$.
\end{theorem}
Proof: Since $E$ is consistent, 
\[
(D\cup\{d_{\vph}\})_E = D_E \cup \{\frac{\vph:\ }{\bot}\}.
\]
Assume that $\vph\notin E$ and that $E$ is an extension of $(D,W)$. Then
\[
E= Cn^{D_E}(W)
\]
and $\vph\notin Cn^{D_E}(W)$. Consequently, 
\[
E= Cn^{D_E}(W) = Cn^{D_E \cup \{\frac{\vph}{\bot}\}}(W)
 = Cn^{(D\cup\{d_{\vph}\})_E} (W).
\]
Hence, $E$ is an extension of $(D\cup\{d_{\vph}\},W)$.

Conversely, assume that $E$ is an extension of $(D\cup
\{d_{\vph}\},W)$. Then,
\[
E=Cn^{(D\cup\{d_{\vph}\})_E} (W)= Cn^{D_E \cup
\{\frac{\vph}{\bot}\}}(W).
\]
Since $E$ is consistent, it follows that $\vph\notin Cn^{D_E}(W)$.
Consequently,
\[
Cn^{D_E}(W) = Cn^{D_E \cup \{\frac{\vph}{\bot}\}}(W) =E.
\]
Hence, $\vph \notin E$ and $E$ is an extension of $(D,W)$.
\hfill$\Box$

We say that a family $\cal F$ of theories
closed under propositional consequence
has a {\em strong system of distinct representatives} (SSDR, for short)
if for every $F\in {\cal F}$ there is a formula $\vph_F\in F$ which does
not belong to any other theory in $\cal F$.

\begin{theorem}\label{th.2}
If $\cal F$ is representable by a default theory and has an SSDR, then
every family ${\cal G} \subseteq {\cal F}$ is representable by a default
theory.
\end{theorem}
Proof: The claim is obvious if ${\cal F}=\{{\cal L}\}$. So, assume that all
members of $\cal F$ are consistent (since $\cal F$ is an antichain,
there are no other possibilities). Let $(D,W)$ be a default theory such
that $Ext(D,W) = {\cal F}$. Define
\[
{\overline D} = D' \cup \{d_{\vph_F}\colon F\in {\cal F}\setminus {\cal
G}\}.
\]
Since all theories in ${\cal F}$ are consistent, the assertion follows
from the definition of an SSDR and from Theorem \ref{th.1}.
\hfill$\Box$

Let us conclude this section with two observations. First,
there are families of theories closed under propositional
consequence which possess SSDRs but which are not representable by 
a default theory (the examples presented in the paper have this
property). Second, not every subfamily of a representable family 
is representable. It follows by the cardinality argument from the the fact 
that there are representable families of cardinality continuum.

\section{Normal default theories}

Our first result in this section describes the family of extensions 
of an arbitrary prerequisite-free normal default theory.

\begin{theorem}\label{ext-norm}
Let $W, \Psi\subseteq {\cal L}$. Let $D = \{\frac{\colon\vph}{\vph}\colon\vph\in \Psi\}$.
If $W$ is inconsistent then $\ext(D,W)=\{{\cal L}\}$. Otherwise,
$\ext(D,W)$ is exactly the family of all theories of the form
$Cn(W\cup \Phi )$, where $\Phi$ is a maximal subset of $\Psi$
such that $W\cup \Phi$ is consistent.
\end{theorem}
Proof: The case of inconsistent $W$ is evident. Hence, let us assume
that
$W$ is consistent. Let $T$ be an extension of $(D,W)$. Since
$W$ is consistent, $T$ is consistent, too. Let $\Phi =
\{\vph\in\Psi\colon
T\not\vdash\neg\varphi\}$. Clearly, $\Phi = c(GD(D,T))$. By (P2),
$T=Cn(W\cup\Phi)$. Moreover, since $T$ is consistent, $W\cup\Phi$ is
consistent.
We will show that $\Phi$ is a maximal subset of $\Psi$ with this
property.
Let $\Phi'$ be such that $\Phi\subseteq\Phi'\subseteq \Psi$. Assume that
$W\cup\Phi'$ is consistent. Then, $T\cup\Phi'$ is consistent.
Hence, $\Phi'\subseteq \Phi$ and, consequently, $\Phi=\Phi'$.

Assume next that $T = Cn(W\cup \Phi )$, where $\Phi$ is a maximal subset
of $\Psi$
such that $W\cup \Phi$ is consistent. Then, it is easy to see that
\[
GD(D,T) = \left\{\frac{\ \colon\vph}{\vph}\colon \vph\in \Phi\right\}.
\]
Hence, $\Phi =c(GD(D,T))$ and $T = Cn(W\cup c(GD(D,T)))$.
Since all defaults in $D$ are prerequisite-free,
it follows by the property  (P2) that $T$ is an extension of $(D,W)$. 
\hfill$\Box$
\ \\

As a corollary, we obtain a full characterization of families of
theories that are representable by normal default theories.

\begin{corollary}\label{32}
A family $\cal T$ of theories in $\cal L$ is representable by a normal
default theory if and only if ${\cal T} = \{ \cal L \}$ or there is a
consistent
set of formulas $W$ and a set of formulas $\Psi$ such that ${\cal T} =
\{ Cn(W \cup \Phi ) : \Phi \subseteq \Psi \ \ \mbox{and}$ $\Phi$ is
maximal so that $W\cup \Phi$ is consistent$\}$.
\end{corollary}
Proof: By Theorem \ref{prereq-free}, $\cal T$ is representable
by a normal default theory if and only if it is representable
by a normal default theory with all defaults  prerequisite-free.
Hence, the assertion follows from Theorem \ref{ext-norm}.
\hfill$\Box$

\begin{corollary}\label{32a}
A family of theories $\cal T$ is representable by a normal default
theory with empty objective part if and only if there is a set of
formulas $\Psi$ such that ${\cal T} = \{ Cn (\Phi ) : \Phi$ is maximal
consistent subset of $\Psi$\}. \hfill$\Box$
\end{corollary}

We will next study the issue of equivalence between normal default
theories. We have already seen that we can replace any normal default
theory with an equivalent
normal prerequisite-free one (Theorem \ref{prereq-free}).
The problem of interest now will be to establish when a normal default
theory can be replaced by an equivalent normal default theory with empty
objective part. We have only partial answers to this problem.

First, consider a normal default theory $(D,W)$ such that $W$ is
inconsistent. Then $\ext(D,W)=\{\cal L\}$. On the other hand, for every
set of normal defaults $D'$, $\ext(D',\emptyset)$ contains only
consistent extensions. Hence, $(D,W)$ is not equivalent to any normal
default theory with empty objective part. From now on we will focus
on normal default theories $(D,W)$ for which $W$ is consistent. 

\begin{theorem}\label{empty-w1}
For every normal default theory $(D,W)$ with $W$ consistent and
finite there exists
a prerequisite-free normal default theory $(D',\emptyset)$ equivalent
to $(D,W)$.
\end{theorem}
Proof: By Theorem \ref{prereq-free}, without loss of generality
we can assume that each
default in $D$ is prerequisite-free.
Define $\omega= \bigwedge W$.

First, assume that the justification of every default
in $D$ is inconsistent with $\omega$. Then, $\ext(D,W)=\{Cn(W)\}$.
Let $D'=\{\frac{\ \colon \omega}{\omega}\}$. Clearly,
$\ext(D',\emptyset) = \ext(D,W)$.

Hence, assume that there are defaults in $D$ whose justifications
are consistent with $\omega$.
For every default $d =\frac{\ \colon\beta}{\beta}$
in $D$, define
$d_\omega = \frac{\ \colon \beta\wedge\omega}{\beta\wedge\omega}$.
Finally, define $D'=\{d_\omega\colon d\in D\}$.
The statement now follows from Theorem \ref{ext-norm} \hfill$\Box$.
\ \\

Next, we will study normal default theories of a special form.
Let $P$ be a subset of the set of atoms of the language $\cal L$.
By a {\em $P$-literal} we mean an element of $P$ or the negation of an
element of $P$. Define 
\[
D^{\comp^P}=
\left\{\frac{\ \colon q}{q}\colon \mbox{where $q$ is a $P$-literal
}\right\}.
\]

XXXXXXXXXXXXXXXXX

A theory $T$ is {\em $P$-complete} if it contains $p$ or $\neg p$ for
every $p\in P$. We will now describe theories representable by default 
theories of the form $(D^{{{\comp^P}}} ,W)$.

\begin{proposition}\label{38}
For every consistent theory $W$, $T$ is an extension of
$(D^{{{\comp^P}}} ,W)$ if and only if a theory $T$ is inclusion-minimal
among theories which are $P$-complete and contain $W$.
\end{proposition}
Proof: If $T$ is an extension of $(D^{{\comp^P}} ,W)$ then $W \subseteq
T$. In addition, since exactly
one of every pair $\frac{\ \colon q}{q}$ and $\frac{\ \colon \neg q}{\neg
q}$ of defaults from $D^{{{\comp^P}}}$ is used
when constructiong $T$, $T$ is $P$-complete. Let $T'\subseteq T$
be $P$-complete and contain $W$. It follows that $T'$ has exactly the
same $P$-literals as $T$. Hence, $T = Cn^{D^{\comp^P}_T}(W) \subseteq
T'$ and, consequently, $T= T'$.

Conversely, let $T$ be inclusion-minimal $P$-complete theory such that
$W\subseteq T$. Let $T_P$ be the set of all $P$-literals in $T$.
Since $W$ is consistent, there exist $P$-complete, consistent theories
containing $W$. Therefore all inclusion-minimal $P$-complete theories
containing $W$ are consistent. Hence, for every atom $p\in P$, exactly
one of $p$ and $\neg p$ belongs to $T_P$.
It is now clear that $D^{\comp^P}_T = \{\frac{\ :\ }{q}\colon q\in T_P\}$.
Hence, $T \supseteq Cn^{D^{\comp^P}_T}(W)$. 
Since $Cn^{D^{\comp^P}_T}(W)$ is $P$-complete and contains $W$, by minimality
of $T$, $T = Cn^{D^{\comp^P}_T}(W)$. Consequently, $T$ is an extension of 
$(D^{\comp^P},W)$. $\hfill\Box$

\begin{proposition}\label{39}
Let $P$ be a subset of the set of atoms of $\cal L$. 
For every consistent $W$, the family ${\cal T}_{W,P}=\{T: T$ is 
inclusion-minimal among all $P$-complete theories containing $W\}$
is representable by a normal prerequisite-free
default theory with empty-objective part.
\end{proposition}
Proof: Proposition \ref{38} implies that we can represent
${\cal T}_W$
by a default theory $(D^{{\comp}} ,W)$. We will now construct  another
set of
defaults $D$ such that $(D,\emptyset )$ represents ${\cal T}_W$.

If the language $\cal L$ has only a finite number of atoms, we can
assume that $
W$
is finite. Hence, the assertion follows from Theorem \ref{empty-w1}.
So assume that $\cal L$ has infinite number of atoms. Let
$p_0,p_1,\ldots$ be an enumeration of atoms in $\cal L$. For the purpose
of this proof we define, for an atom $p$, $0p = p$, $1p = \neg p$.

Consider a set $\tree^W$ of all finite sequences
$ \la \epsilon_0p_0,\ldots ,\epsilon_np_n\ra$ such that
$W \cup \{\epsilon_0p_0,\ldots ,\epsilon_np_n\}$ is consistent.

The set $\tree^W$ forms a {\em tree}. That is, if $ \la
\epsilon_0p_0,\ldots ,\epsilon_np_n\ra$ belongs to $\tree^W$ and $m < n$
then also $ \la \epsilon_0p_0,\ldots ,\epsilon_mp_m\ra$ belongs to
$\tree^W$.

An infinite branch  in $\tree^W$ determines the infinite
sequence $\la \epsilon_0p_0,\ldots \ra$. Since for
all $n$, $W \cup \{\epsilon_0p_0,\ldots ,\epsilon_np_n\}$ is consistent,
$Cn (\{ \epsilon_0p_0,\ldots ,\epsilon_np_n,\ldots \})$ is also consistent. 
It is also complete and therefore, as $W$ is
consistent with it, $W \subseteq Cn( \{ \epsilon_0p_0,\ldots
\})$. Conversely, if  $T$ is
consistent and complete then there is  a sequence
$\la \epsilon_0p_0,\ldots \ra$ such that
$T = Cn (\{ \epsilon_0p_0,\ldots \} )$. If
$W\subseteq T$ then for all $n$, $W \cup \{\epsilon_0p_0,\ldots
,\epsilon_np_n\}$ is consistent. Thus we proved that there is a
one-to-one correspondence between the branches in
$\tree^W$ and complete, consistent theories containing $W$.

Now define:
\[
D =\left\{ \frac{\ : \epsilon_0p_0\wedge\ldots
\wedge\epsilon_np_n}{\epsilon_0p_0
\land\ldots \land\epsilon_np_n} : \langle\epsilon_0p_0,\ldots
,\epsilon_np_n\ra
\in \tree^W\right\}
\]
We show that the extensions of $(D,\emptyset )$ are precisely the
theories
of the form $Cn (\{ \epsilon_0p_0,\ldots ,\epsilon_np_n,\ldots \} )$,
where $\la \epsilon_0p_0,\ldots ,\epsilon_np_n,\ldots \ra$ is an
infinite branch through $\tree^W$.

Indeed, if $\la \epsilon_0p_0,\ldots ,\epsilon_np_n,\ldots \ra$ is an
infinite branch through $\tree^W$ then $T = Cn (\{ \epsilon_0p_0,\ldots
,\epsilon_np_n,\ldots \} )$ is a complete theory. The only default rules
in $D$ that have conclusions in $T$ are the rules
$ \frac{\ : \epsilon_0p_0\wedge\ldots \wedge\epsilon_np_n}{\epsilon_0p_0
\land\ldots \land\epsilon_np_n}$ for $n \in N$. This implies that
$Cn (\{ \epsilon_0p_0,\ldots ,\epsilon_np_n,\ldots \} )$ is an extension
of $(D, \emptyset )$.

Conversely, if $T$ is an extension of $(D,\emptyset)$ then if
$ \frac{\ : \epsilon_0p_0\wedge\ldots \wedge\epsilon_np_n}{\epsilon_0p_0
\land\ldots \land\epsilon_np_n}$ is a generating rule for $T$ then for
all $m < n$, $\frac{\ : \epsilon_0p_0\wedge\ldots\wedge
\epsilon_np_m}{\epsilon_0p_0 \land\ldots \land\epsilon_np_m}$ is
also a generating rule for $T$. Next, since $W = \emptyset$, $T$ must be
consistent. This means that if two rules $ \frac{\ :
\epsilon_0p_0\wedge\ldots \wedge\epsilon_np_n}{\epsilon_0p_0
\land\ldots \land\epsilon_np_n}$ and $ \frac{\ :
\delta_0p_0\wedge\ldots \wedge\delta_mp_m}{\delta_0p_0
\land\ldots \land\delta_mp_m}$  are both generating for $T$ then  $m
\leq n$ and
 $\delta_0 = \epsilon_0, \ldots \delta_m =
\epsilon_m$ or $n \leq m$ and $\delta_0 = \epsilon_0, \ldots \delta_n =
\epsilon_n$. Thus, in order to complete our argument it is enough to
show that the set of generating rules for $T$ is infinite. Assume
otherwise.
Then,
there exists a sequence $\la \epsilon_0p_0,\ldots ,\epsilon_np_n
\ra$ such that $T = Cn(\{ \epsilon_0p_0,\ldots ,\epsilon_np_n \} )$.
But recall that $\la \epsilon_0p_0,\ldots ,\epsilon_np_n
\ra \in \tree^W$. Therefore $W \cup \{ \epsilon_0p_0,\ldots
,\epsilon_np_n \}$ is consistent and so there is a complete theory $T'$
containing $W \cup \{ \epsilon_0p_0,\ldots ,\epsilon_np_n \}$. In
particular $p_{n+1} \in T'$ or $\neg p_{n+1} \in T'$. Without loss of
generality assume that $p_{n+1}$ (i.e. $0p_{n+1}$) belongs to $T$.
Setting $\epsilon_{n+1} = 0$ we have
\[
\la  \epsilon_0p_0,\ldots ,\epsilon_{n+1}p_{n+1}\ra \in \tree^W.
\]
Hence, the default rule $\frac{\ :
\epsilon_0p_0\wedge\ldots \wedge\epsilon_{n+1}p_{n+1}}{\epsilon_0p_0
\land\ldots \land\epsilon_{n+1}p_{n+1}}$  belongs to $D$. Since
$T \subseteq T'$, $p_{n+1}$ is consistent with $T$. Therefore
$\epsilon_0p_0 \land\ldots \land\epsilon_{n+1}p_{n+1}$ is consistent
with $T$ and thus $\epsilon_0p_0 \land\ldots \land\epsilon_{n+1}p_{n+1}$
belongs to $T$. In particular $p_{n+1} \in T$, a contradiction.
Therefore the set of generating default rules for $T$ is infinite and
determines a branch through $\tree^W$. Thus $T$ is a complete theory
containing $W$. This completes the proof. $\hfill\Box$

\section{Representability with Closed World Assumption}

Next, we explore the connections of normal default logic with the Closed 
World Assumption.
%
Consider a set of atoms $P$. Define the set of defaults
\[
D^{{\cwa}^P} = \left\{\frac{\ \colon \neg p}{\neg p}\colon p\in P
\right\}.
\]
Informally, a default $\frac{\ \colon \neg p}{\neg p}$ allows 
us to derive $\neg p$ if $p$ is not derivable. This has the flavor of
the Closed World Assumption. The exact connection
with $\Cwa$ is given by the following result
\cite{mt93}: If $P=\at$ then $W$ is $\Cwa$-consistent if and only if
$(D^{{\cwa}^P},W)$ possesses a unique consistent extension.

\begin{theorem} \label{cwa-p}
For every normal default theory $(D,W)$ in $\cal L$ there exists a
language ${\cal
L}'\supseteq {\cal L}$, a set of atoms $P$ in ${\cal L}'$, and
$W'\subseteq {\cal L}'$ such that $(D,W)$ is semi-equivalent to
a default theory $(D^{{\cwa}^P},W')$.
\end{theorem}
Proof: By Theorem \ref{prereq-free} we can assume that all
defaults in $D$ are prerequisite-free. Let $\Psi$ be the set of
consequents
of defaults in $D$. For each $\psi\in \Psi$ select a new atom not
belonging
to $At$ (recall that $\at$ is the set of atoms in $\cal L$). This atom is
denoted
by $p_\psi$ and the set $P$ is defined as
$\{p_\psi : \psi\in\Psi\}$. Define now ${\cal L}'$ to be the language
generated
by
the set of atoms $At' = At
 \cup P$. Next, define $V$ as this set of formulas:
\[
\{ \neg p_\psi \Leftrightarrow \psi\colon \psi\in \Psi\}.
\]
We notice the following fact:
\begin{description}
\item[(F1)] Let $\Phi\subseteq \Psi$. Then $W\cup \Phi$ is consistent 
if and only if
$W\cup V \cup \{ \neg p_\psi : \psi \in \Phi\}$ is consistent.
\end{description}
Indeed, for a model  $v$ of $W\cup \Phi$, define $v'$ as
follows:
\[
v'(p) =\left \{ \begin{array}{ll}
v(p) & \mbox{if $p\in At$}\\
1 - v(\psi ) & \mbox{if $p = p_\psi$}
\end{array}
\right.
\]
It is clear that $v'$ is a model of $W\cup V \cup \{\neg p_\psi :
\psi\in \Phi\}$. Conversely, when $v'$, a valuation of $At'$ is a model
of
$W\cup V \cup \{\neg p_\psi : \psi\in \Phi\}$  then $v = v'|_{At}$ is a
model of
$W \cup \{ \psi : \psi\in \Phi\}$. Hence, (F1) follows.

Observation (F1) implies that $\Phi$ is a maximal subset of $\Psi$ consistent
with $W$ if and only if $\{\neg p_\psi :\psi\in \Phi\}$ is a maximal
subset of
$\{\neg p_\psi : \psi \in \Phi\}$ which is consistent with $W\cup V$.

Next, observe that if $\Phi$ is a maximal set of formulas contained in
$\Psi$ and consistent with
$W$ then for all $\theta\in \Psi\setminus\Phi$
\[
W\cup V\cup\{ \neg p_\psi : \psi \in \Phi\} \vdash p_\theta .
\]
We are now ready to construct the desired default theory. We put
$W' = W \cup V$ and $D' = \{ \frac{\ :\neg p_\psi}{\neg p_\psi} : \psi\in
\Psi\}$. Clearly, $D' = \Cwa^P$. 
Using the observations listed above, we will now show that
the theory $(D',W')$ semi-represents $\ext(D,W)$. 

We will apply Theorem 4.21 of
\cite{mt93}. In order to apply this result, we need to show that whenever
$\Phi\subseteq \Psi$ is a maximal subset of $\Psi$ consistent with $W$
then
$ T = Cn(W\cup V \cup \{ \neg p_\psi : \psi \in\Phi\} )$ is
inclusion-minimal
and $\sqsubseteq_P$-minimal among theories that are complete with
respect to
$P$. In addition, we need to show that, $Cn (W\cup \Phi ) = T\cap {\cal
L}$.

To this end, we proceed as follows.
When $\Phi$ is a maximal subset of $\Psi$ consistent with $W$ then, by
the above fact, $\{ \neg p_\psi :\psi\in \Phi\}$ is a maximal subset
of $\{\neg p_\psi : \psi\in \Phi\}$ consistent with $W\cup V$. Thus
$\{ p_\phi : \phi \in \Psi\setminus\Phi\}$ is a minimal set of atoms in
a theory $T$ which contains $W\cup V$. Moreover, among such theories
$Cn(W\cup V\cup \{\neg p_\psi : \psi\in \Phi\})$ is clearly
inclusion-minimal.
Moreover, by the argument used in the proof of (F1) we have:
\[
Cn(W\cup V\cup \{ \neg p_\psi : \psi \in \Psi \} ) \cap {\cal L} =
Cn(W \cup \Phi )
\]
Thus the extensions of $(D,W)$ are restrictions to $\cal L$ of
extensions
of $(D',W')$. It is easy to see that such extension of $(D',W')$ is
unique.

In order to apply Theorem 4.21 we also need to prove the converse
statement, that is, that if $T$ is complete
with respect to $P$, $T$ contains $W\cup V$, $T$ is
$\sqsubseteq$-minimal
with these properties, and $\Phi_T = \{\psi : \psi \in \Psi\} \cap T$,
then $\Phi_T$ is consistent with $W$ and is maximal among the subsets of
$\Phi$ with this property, $T = Cn(W\cup V = \{ \neg p_\psi :
\psi \in \Phi_T\} )$ and $Cn(W \cup \Phi_T) = T\cap {\cal L}$.
Since the argument is entirely analogous to the one used above,
we omit the details. $\hfill\Box$
\ \\

\section{Conclusions}

The concepts studied in this paper, representability and equivalence,
are of key importance for default logic and its applications.
Representability provides insights into the expressive power of default
logic, while equivalence provides normal form results for default logic,
allowing the user to find simpler representations for his/her default
theories.

In this paper we characterized those
families of theories that can be represented by default theories with 
a finite set of defaults (Theorem \ref{th-1} and Corollary \ref{th-2}).
However, we have not found a characterization of families of theories 
that are representable by default theories with an infinite set of
defaults. This problem seems to be much more difficult and remains open. 
In the paper, we present two countable families that are not
representable and completely solved the representability problem if
inifintary defaults are allowed.

We also studied representability by means of normal default theories.
Here, our results are complete. Corollary \ref{32} provides a full
description of families of theories that are collections of extensions
of normal default theories. 

Another notion studied in the paper was equivalence of default theories.
We showed (Theorem \ref{prereq-free}) that for every normal default 
theory there exists a normal default theory consisting of prerequisite-free 
defaults and having exactly the same extensions as the original one. We also
exhibited some cases when, for a given normal default theory, an equivalent 
normal default theory can be found with empty objective part. Finding 
a complete description of normal default theories for which it is possible 
remains an open problem.

\section*{Acknowledgements}

The first and third authors were partially supported by the  NSF grant 
IRI-9400568. The work of the second author was partially 
supported by the ESPRIT III
Basic Research Project 6156 DRUMS II on Defeasible Reasoning
and Uncertainty Management Systems.

\singlespace

\end{document}